# Etching and Narrowing of Graphene from the Edges

Xinran Wang and Hongjie Dai[*]

*Department of Chemistry and Laboratory for Advanced Materials, Stanford University, Stanford, CA 94305, USA*

*\* Correspondence to hdai@stanford.edu*

**Abstract:**

Large scale graphene electronics desires lithographic patterning of narrow graphene nanoribbons (GNRs) for device integration. However, conventional lithography can only reliably pattern ~20nm wide GNR arrays limited by lithography resolution, while sub-5nm GNRs are desirable for high on/off ratio field-effect transistors (FETs) at room temperature. Here, we devised a gas phase chemical approach to etch graphene from the edges without damaging its basal plane. The reaction involved high temperature oxidation of graphene in a slightly reducing environment to afford controlled etch rate ($\leq$ ~1nm/min). We fabricated ~20-30nm wide GNR arrays lithographically, and used the gas phase etching chemistry to narrow the ribbons down to <10nm. For the first time, high on/off ratio up to ~$10^4$ was achieved at room temperature for FETs built with sub-5nm wide GNR semiconductors derived from lithographic patterning and narrowing. Our controlled etching method opens up a chemical way to control the size of various graphene nano-structures beyond the capability of top-down lithography.

Graphene has attracted much attention as a novel two-dimensional system with high potential for future electronics.[1,2] Narrow graphene nanoribbons (GNRs) have been demonstrated as semiconducting wires to afford field-effect transistors (FETs) with high on/off ratios at room temperature.[3-9] Among various methods of producing narrow GNRs ranging from chemical sonication[3] to unzipping of carbon nanotubes,[5,10] top-down lithographic patterning of large pristine graphene sheet into GNRs is appealing for large-scale device integration.[2] Thus far, patterning methods have only produced GNR arrays down to ~20nm in width (except for short, narrow constrictions[9]) limited by lithography resolution,[7,8] while GNRs narrower than ~5nm with sufficient bandgaps are needed for room temperature FET operation.[3-6,11-13] Here, we devise a gas phase chemical approach to etch graphene from the edges without damaging the basal plane of graphene. The reaction involves high temperature oxidation of graphene in a slight reducing environment, to afford controlled ($\leq$ ~1nm/min) etching of graphene. We fabricated ~20-30nm wide GNR arrays by electron beam lithography, and subsequently narrowed the GNRs by the gas phase etching chemistry. As the GNRs were narrowed to the $\leq$ ~5nm regime, bandgap opening related to quantum confinement was clearly observed in GNR-FETs. For the first time, high on/off ratio up to ~$10^4$ was achieved at room temperature for devices of ~5nm wide GNRs derived from lithographic patterning. Parallel arrays of ~8nm wide GNRs were used to produce FETs with on/off ratio ~50 and on-currents far exceeding those of single-ribbon devices. Controlled chemical etching could play important roles in tailoring the dimensions of graphene for large scale device integration.

To devise a gas-phase chemistry for etching and narrowing graphene from the edges without creating defects in the basal plane, we investigated chemical etching of few-layer ($\leq$ 3-layer) mechanically exfoliated graphene on 300nm $SiO_2$/Si substrate under various oxidation conditions at high temperatures (Fig. 1, supplementary Fig. S1&S2. Also see Methods for experimental details). We varied the partial pressure of $O_2$, and introduced Ar, $H_2$ or $NH_3$ as dilution gas or to provide a reducing environment. We found that the etching of graphene was highly dependent on the gas mixture, reaction temperature as well as the number of layers of graphene (Table S1). At 750°C, ~100mTorr of $O_2$ diluted by ~1Torr of Ar or $H_2$ gave an etch rate of ~3-5nm/min (~6-7nm/min) for single (double) layer graphene (Fig. S1). Interestingly, when using ~100mTorr $O_2$ with ~1Torr 10% $NH_3$ in Ar as additive, we observed a considerably slower etching rate of ~2-2.5nm/min (~4nm/min) for single (double) layer graphene (Fig. S2). We then reduced the $O_2$ partial pressure to ~25mTorr aimed at slower etching. Under ~25mTorr $O_2$ and ~1Torr $H_2$, the etching rate at 800°C was ~3-3.5nm/min (~3.8-5nm/min) for single (double) layer graphene (Fig. S1). By introducing $NH_3$ into the system, we found that the etching rate was further decreased to less than ~1nm/min under ~25mTorr of $O_2$ in ~1Torr of 10% $NH_3$ in Ar (Fig.1, S1 and Table S1). The slow etching rate was highly desirable for controlled etching and narrowing of graphene and GNRs down to the $\leq$ ~5nm regime.

We observed that graphene etching occurred mostly at the edges and proceeded inward isotropically (Fig. 1, S1&S2). Due to bond disorders and functional groups, the chemical reactivity of graphene edge carbon atoms was higher than the perfectly bonded sp$^2$ carbon atoms in the basal plane.[14] As a result, graphene sheets were etched uniformly from the edges under controlled oxidation conditions. Occasionally, we observed holes or trenches formed in graphene after reactions, due to etching initiated at point defects or line defects within graphene plane respectively (Fig.1d-f & Fig.S1).[14-16] Certain line defects were visible under atomic force microscopy (AFM) such as wrinkles (Fig. 1d), though not all defects were observed by AFM (Fig.S1a-b&d-e). Importantly, Raman spectroscopic mapping and imaging revealed that etching from the edges did not damage the basal plane of pristine graphene, as indicated by the absence of defect related Raman band (D band) in the plane of graphene sheets after etching (e.g. Fig.1g-i & Fig.S2c-d). We observed red-shifts of the Raman G band of graphene after etching in both $O_2/NH_3$ (Fig. 1i) and $O_2/Ar$ conditions, which was not fully understood but could be due to strain in the graphene sheets as indicated by the formation of wrinkles sometimes observed after the high temperature reaction (Fig. 1e, Fig. S1b,S1e).[17,18]

Our graphene etching process involves $O_2$ oxidation of graphene into CO or $CO_2$. During etching, $O_2$ molecules exothermally dissociate and form bonds with dangling carbon atoms on defect sites and edges.[15,16] The slower etching rate under $NH_3$ environment was attributed to in-situ $NH_3$ reduction of oxygen groups formed during graphene oxidation. Our recent work on the reduction of graphene oxide suggested that $NH_3$ was more effective than $H_2$ in reducing oxygen groups in graphene oxide.[19] The $NH_3$ reduction effect could impede the oxidation of graphene when mixed with oxygen. Also interesting was that under the same reaction condition, we found that thicker graphene sheets were etched faster than thinner ones (Table S1, Fig. S2), similar to earlier observations made on thick graphite.[15,16] This was not understood but could be due to synergistic effects of oxygen groups at the edges of adjacent layers, giving self-catalyzed etching at the edges of multi-layer graphene.[16] As a control, we found that heating graphene in pure ammonia did not give etching effects (Fig.S3a).

Next, we fabricated GNR arrays by lithographic patterning, and used our gas phase reaction to narrow the as-made ribbons down to several nanometres in width. GNR arrays were fabricated by electron beam lithography (EBL) and Ar plasma etching on 300nm $SiO_2/Si$ substrate (Fig. 2a, see Methods for detailed fabrication process). Unlike previously used electron beam resist as etch mask,[7-9] we used thin (~6.5nm) Al lines (defined by single-pixel electron beam) on exfoliated graphene as etch mask (Fig. 2a and supplementary information). The widths of GNRs were measured by AFM with finite tip size deconvolution.[3] The as-patterned GNRs were down to ~20nm in width, with a mean edge roughness $\delta w = \left( w_{max} - w_{min} \right)/2 \leq$ ~5nm (Fig. 2b-g). The edge roughness was caused by random fluctuating factors in lithography and plasma etching processes.[6-9] Lithographic patterning using Al as etch mask is versatile in patterning graphene into various structures with high reproducibility and consistency. For examples, we readily fabricated $w$~20nm GNR alphabetic characters and junctions (Fig. 2d-g), which could be interesting structures to study electron transport along various crystallographic directions of graphene.[20,21]

To afford GNR-FETs with high $I_{on}/I_{off}$ ratio at room temperature, $w \leq$ ~5nm GNRs with sufficient bandgap ($E_g >> k_B T$~26meV) are needed.[3-6,11-13] We narrowed down the as-made $w$~20nm GNRs using the 0.5-1nm/min etching condition (25mTorr $O_2$ in 1Torr $NH_3$/Ar at 800°C). It was difficult to narrow GNRs with good width control in several control experiments under other conditions (Fig. S3). We succeeded in narrowing the GNRs uniformly from ~20nm to ~8nm without obvious breaks along the ribbons (Fig. 3a&b). However, further narrowing typically resulted in breaks due to edge roughness and width variations in as-patterned GNRs. We obtained $w \leq$ ~5nm ribbons by over-etching, with most GNRs evolving into discontinuous segments down to sub-5nm in width. Some of the segments exhibited sufficiently long length useful for integration into FET devices for electrical measurements (Fig. 4d). Edge roughness of as-made GNRs by lithographic patterning is currently a limiting factor in producing long, uniform ultra-narrow GNRs over large areas.

We carried out Raman spectroscopic measurements on the as-made and narrowed GNRs (Fig. 3c&d). For $w$~20nm as-made GNRs, several pronounced peaks were observed, including D band (~1340cm$^{-1}$), G band (~1590cm$^{-1}$), D' band (~1620cm$^{-1}$) and 2D band (~2670cm$^{-1}$) (Fig. 3c). The intensity ratio between D and G bands $I_D/I_G$ was ~1-2. The presence of defect-related D and D' bands was attributed to the edges of GNRs since no D band was observed in the parent graphene sheet.[5,22] For a $w$~20nm GNR, about 1% carbon atoms are at the edges, resulting in D and D' bands as expected.[22] We observed lower $I_D/I_G$ ratio in wider GNRs (supplementary information and Fig. S7), consistent with reduced edge effects. After narrowing below ~10nm, the intensity of G and 2D bands of graphene reduced, $I_D/I_G$ ratio increased and the G and D' peaks were broadened to form a single and up-shifted G peak (Fig.3d).[22] These changes in Raman spectra were due to higher percentage of edge atoms in narrowed GNRs. The $I_D/I_G$ ratios of our narrowed GNR were larger than similar width GNRs derived chemically,[4] suggesting higher degree of edge roughness and disorder in the former.

We fabricated electrical devices on as-made and narrowed GNRs with Pd source and drain electrodes and heavily doped Si back-gate (see Methods). At room temperature, devices of as-made $w$~20nm GNR and GNR arrays typically showed $I_{max}/I_{min}$ current ratios less than ~3 (Fig. S4),[5,6,8] indicating insufficient bandgap (compared to $k_BT$~26meV) for room temperature FETs. The bandgap of a perfect GNR is predicted to scale inverse linearly with GNR widths, $E_g$~$(0.3\text{-}1.5eV{\cdot}nm)/w(nm)$ depending on the orientation and edge configuration of the ribbons.[12,13] For GNRs narrowed down to ~10nm, we observed an $I_{max}/I_{min}$ ratio ~ 10 at room temperature (Fig. S5). With long GNRs narrowed down to $w$~8nm with good continuity (Fig. 3b), we fabricated a FET using an array of GNRs in parallel and inter-digitized electrodes as contacts (Fig. 4a). We observed high $I_{on}/I_{off}$ ratio up to ~50 at room temperature and on-state current of ~20μA (~40 times that of single ribbon devices) at $V_{ds}$= -500mV (Fig. 4b&c) with ~40 GNR sections (channel length ~ 160nm).

For GNRs narrowed down to the $w \leq$ ~5nm regime, limited by discontinuity in the ribbons, we fabricated devices only on single ribbons rather than with parallel arrays. Figure 4e&f show room temperature $I_{ds}$-$V_{gs}$ and $I_{ds}$-$V_{ds}$ characteristics of a GNR-FET with a $w$~4nm GNR (see Fig. S6 for another example). The device showed ambipolar transport in air with $I_{on}/I_{off}$ > $10^4$ (Fig. 4e), a clear evidence for bandgap opening by lateral quantum confinement.[3-6,11-13] This was the highest room-temperature $I_{on}/I_{off}$ ratio reported for GNRs derived from lithographic patterning.[6-9] In vacuum, the device show an intrinsic n-type behaviour with threshold voltage shifted to the negative gate-voltage side due to desorption of physisorbed species including oxygen.[23] Based on the ambipolar $I_{ds}$-$V_{gs}$ characteristics, we estimated the bandgap of the GNR to be $E_g$ ~ 0.4eV from $I_{on}/I_{off}$ ~ $exp(E_g/2k_BT)$ since the off-state current was thermally activated over a body-Schottky Barrier of ~$E_g/2$ (Ref. 8).

We developed controlled chemical narrowing of graphene to afford quantum confined structures. Parallel GNR arrays were made to afford graphene FETs with high on-currents and high on/off ratio of ~50 at room temperature. Single ribbons based on lithography were narrowed below ~5nm to afford on/off ratios of ~$10^4$. On the single-ribbon basis, our narrowed GNRs afforded lower on-currents than our previously reported chemically derived sub-10nm GNRs on the same 300nm $SiO_2$ substrate.[3] The narrowed lithography-derived GNRs showed rougher and more disordered edges, as reflected from Raman spectroscopy data. Edge disorder caused scattering effects that contributed to the low on-currents of GNR devices.[4,24-26] There are variations between devices such as the doping, likely due to the differences in the detailed edge structures. Clearly, the success of our approach is currently limited by the edge roughness introduced in the patterning process, and much effort should be directed in the future towards making smooth edges by either improving lithographic patterning processes, or developing novel chemical means to perfect the edges. Some recent experiments have shown the promise of making graphene and GNRs with atomically well defined edges by anisotropic etching,[27-30] which could be combined with narrowing to produce long, uniform sub-5nm GNR semiconductors to produce high performance graphene transistors for potential digital applications. This could be an appealing roadmap for graphene especially since chemical narrowing of GNRs with well defined crystallographic orientations should be possible to afford semiconducting GNRs with desired edge structures.

**Methods**

**Preparation of mechanically exfoliated graphene, microscopy and spectroscopy characterizations and lithographic patterning of GNRs**

Graphene sheets used in etching experiments under various conditions were mechanically exfoliated from highly oriented pyrolitic graphite crystals on a thermally grown ~300nm $SiO_2/p^{++}$ Si substrate using Scotch® tape.[31] A subsequent annealing in ~2Torr $H_2$ at 800°C for ~15mins was done to clean the substrate and graphene sheets. Then AFM and Raman mapping were used to characterize the graphene sheets before and after the gas phase reaction. Raman mapping was done by a Horiba Jobin Yvon LabRAM HR Raman microscope with a 633nm He-Ne laser excitation (spot size ~1μm, power ~5mW). We used 100nm step size and 4 seconds accumulation time to map the graphene sheets.

Graphene sheets used for patterning GNRs were mechanically exfoliated on a thermally grown ~300nm $SiO_2/p^{++}$ Si substrate with pre-patterned Ti/Au markers and were annealed at ~600°C for ~15mins (Ti/Au markers melt at higher temperature). We used optical microscope to locate few-layer (≤ 3-layer) graphene sheets and then did Raman spectroscopy to determine the number of layers. We spun ~70nm PMMA with molecular weight of 950kDa as EBL resist. Single pixel lines were exposed in a Raith 150 system in the Stanford Nanofabrication Facility with acceleration voltage of 10keV and line dosage of 650μC/cm. Wider GNRs could be made by exposing areas instead of single pixel lines. The development was done in cold (4˚C) 1:3 methyl iso-butyl ketone: isopropanol solution for 75 seconds to afford good edge roughness in the resist profile.[32] A 6.5nm thick Al film was then electron beam evaporated, followed by standard liftoff. The Ar plasma etching was done in a MRC plasma etcher for ~20-30s (depending on number of layers of graphene), under an Ar flow rate of ~10cm³/min, chamber pressure ~40mTorr and plasma power ~10W. After plasma etching, the chips were soaked in 0.1mol/L KOH water solution for ~2minutes to remove the Al lines. We then annealed the chips in ~2Torr $H_2$ at 600˚C to clean resist residues from the substrate.

**Gas phase chemical etching of graphene in a vacuum furnace**

Gas phase etching of graphene sheets and narrowing of GNRs were carried out in a vacuum furnace connected to a mechanical pump, with a base pressure of ~15mTorr. Note that a leak-free vacuum system was essential to high reproducibility of etching results. According to our experience, the etching rate is sensitive to the detailed configuration of the vacuum furnace and may vary in different systems. The vacuum level in the furnace was monitored by a Millipore CML series 0-100Torr vacuum gauge. We used Praxair ultra high purity grade Ar, $H_2$ and 10% $NH_3$ in Ar and research grade $O_2$ in our experiments. To adjust the pressure of the gases, we first close all the gas valves and record the base pressure of the system. Then we opened one gas at a time and adjusted the pressure by manual valves. After the pressure reached the target and remained stable for ~1minute, we closed the valve and moved on to the next gas. Finally we opened all the gas valves and started heating to a desired temperature. Since we adjusted the pressure manually, there might be some slight pressure variations for different batches, which could cause slight variations in etching rates.

**Fabrication and electrical characterization of GNR devices**

After making (or narrowing) GNRs on the ~300nm $SiO_2$/$p^{++}$ Si substrate with pre-patterned Ti/Au markers, we used AFM to locate the GNR or GNR array relative to a marker. We then used EBL to define source and drain, followed by ~20nm Pd evaporation and liftoff. The devices were annealed in Ar at ~200°C to anneal metal contacts. The electrical data of GNR devices were taken by a standard semiconductor analyzer (Agilent 4156C) inside a Lakeshore table-top cryogenic vacuum probe station connected to a turbo pump. The base pressure of the system was ~$10^{-6}$ Torr.

**Acknowledgement**

This work was supported by Intel, MARCO MSD center, Graphene MURI supported by Office of Naval Research. We gratefully thank James W. Conway in Stanford Nanofabrication Facility for helpful discussions.

**Author contributions**

X. W. and H. D. conceived and designed the experiments. X. W. performed the experiments and analyzed the data. X. W. and H. D. co-wrote the paper. All authors discussed the results and commented on the manuscript.

**Figure captions:**

**Figure 1.** Gas phase chemical etching and narrowing of graphene sheets. The etching condition used was ~25mTorr of $O_2$ in the presence of ~1Torr of 10% $NH_3$ in Ar at 800°C for 1 hour. (a) & (b) show the same graphene sheet imaged by AFM as made (a) and after etching (b). (c) shows the overlay images before (yellow) and after (red) etching, with uniform etching from the edges observed at an etching rate of ~0.5nm/min. (d)-(f) show another set of graphene sheet etching data. A wrinkle with high strain (bright strip) on the as-made graphene was etched away to form a trench. Etching rate was ~0.8nm/min. Note that new wrinkles (bright lines in (e)) were observed after the reaction, probably caused by thermal effects. (g) & (h) Raman G band mapping of the same graphene sheet as in (d) and (e) before and after etching, respectively. (i) Averaged Raman spectra from the graphene plane in (g) and (h). The 2D band can be fit into 4 Lorentzians for both as-made and after etching cases, indicating 2-layer graphene for both before and after etching.[21] No obvious D band was observed after etching in the plane of graphene.

**Figure 2.** Lithographically patterned GNR arrays and junctions. (a) Schematics of the fabrications process. (b) An AFM image of a *w*~20nm GNR array at ~200nm pitch. (c) A high resolution AFM image of a *w*~22nm GNR array at ~500nm pitch. (d)-(g) AFM images of various GNR structures including alphabetic characters and zigzag junctions.

**Figure 3.** Gas phase chemical narrowing of GNRs. (a) & (b) AFM images of the same GNR array as made (a) and after chemical narrowing (b), respectively. The narrowing condition was ~25mTorr of $O_2$ in the presence of ~1Torr of 10% $NH_3$ in Ar at 800°C for 10mins. The GNRs were narrowed uniformly from ~20nm to ~8nm with no obvious breaks. The etching rate was ~0.6nm/min. (c) & (d) Averaged Raman spectra of the same GNR array in (a) and (b), respectively.

**Figure 4.** Field-effect transistors from lithographically patterned and chemically narrowed GNRs and parallel GNR arrays. (a) AFM images of a *w*~8nm GNR array FET (top panel, the same array as in Fig. 3b) with inter-digitized electrodes to contact ~40 GNRs in parallel (channel length ~ 160nm). Bottom panel is a zoom-in image of the device. Source (S) and drain (D) contacts are marked on both images. (b) Room temperature $I_{ds}$-$V_{gs}$ characteristics of the GNR array FET in (a) probed in air. The $I_{on}$/$I_{off}$ ratio was ~50. (c) $I_{ds}$-$V_{ds}$ characteristics of the same device in (a) probed in air. From top to bottom, $V_{gs}$= -40V to 50V in 10V steps. (d) AFM of an as-made and narrowed GNR with breaks along the ribbon. The highlighted part was sub-5nm in width, which was used to make a GNR-FET. (e) Room temperature $I_{ds}$-$V_{gs}$ characteristics of the GNR-FET fabricated on the GNR in (d). The device showed an $I_{on}$/$I_{off}$ ratio of higher than $10^4$ probed in air and in vacuum. $V_{ds}$=10mV for both curves. (f) $I_{ds}$-$V_{ds}$ characteristics of the same device in (e) probed in vacuum. From top to bottom, $V_{gs}$= 40V to -30V in -10V steps. Inset shows the AFM image of the GNR-FET.

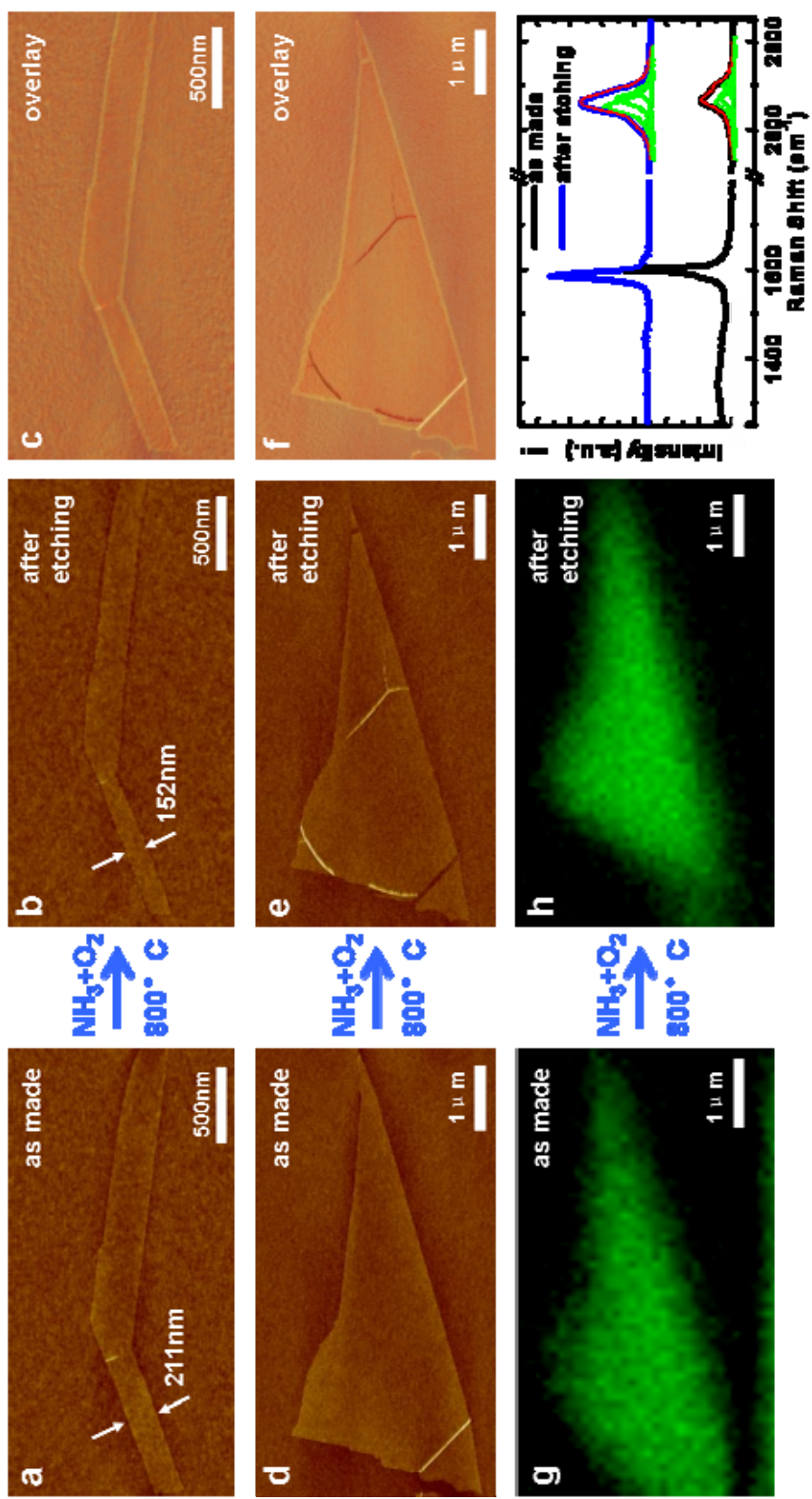

Figure 1

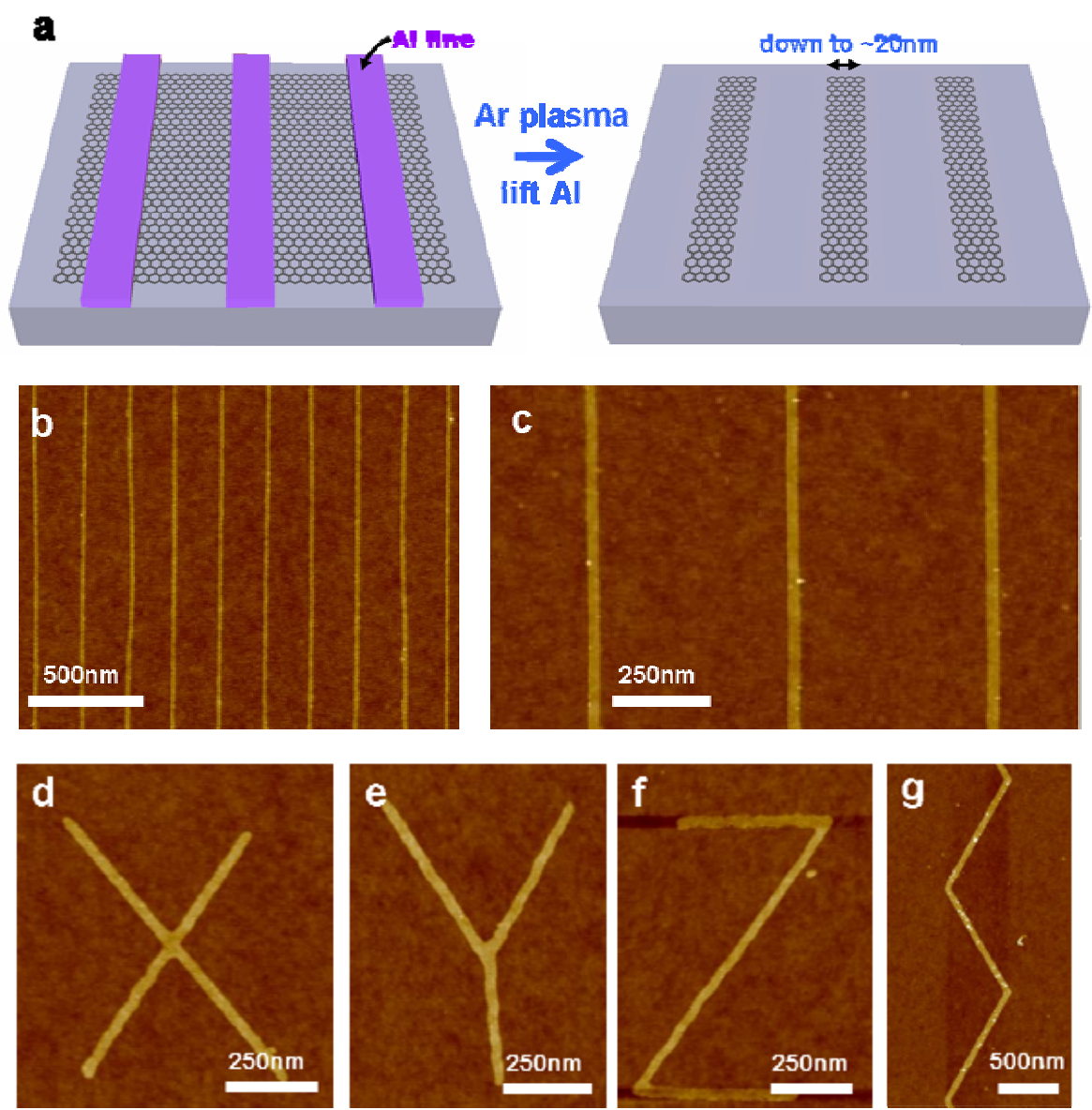

Figure 2

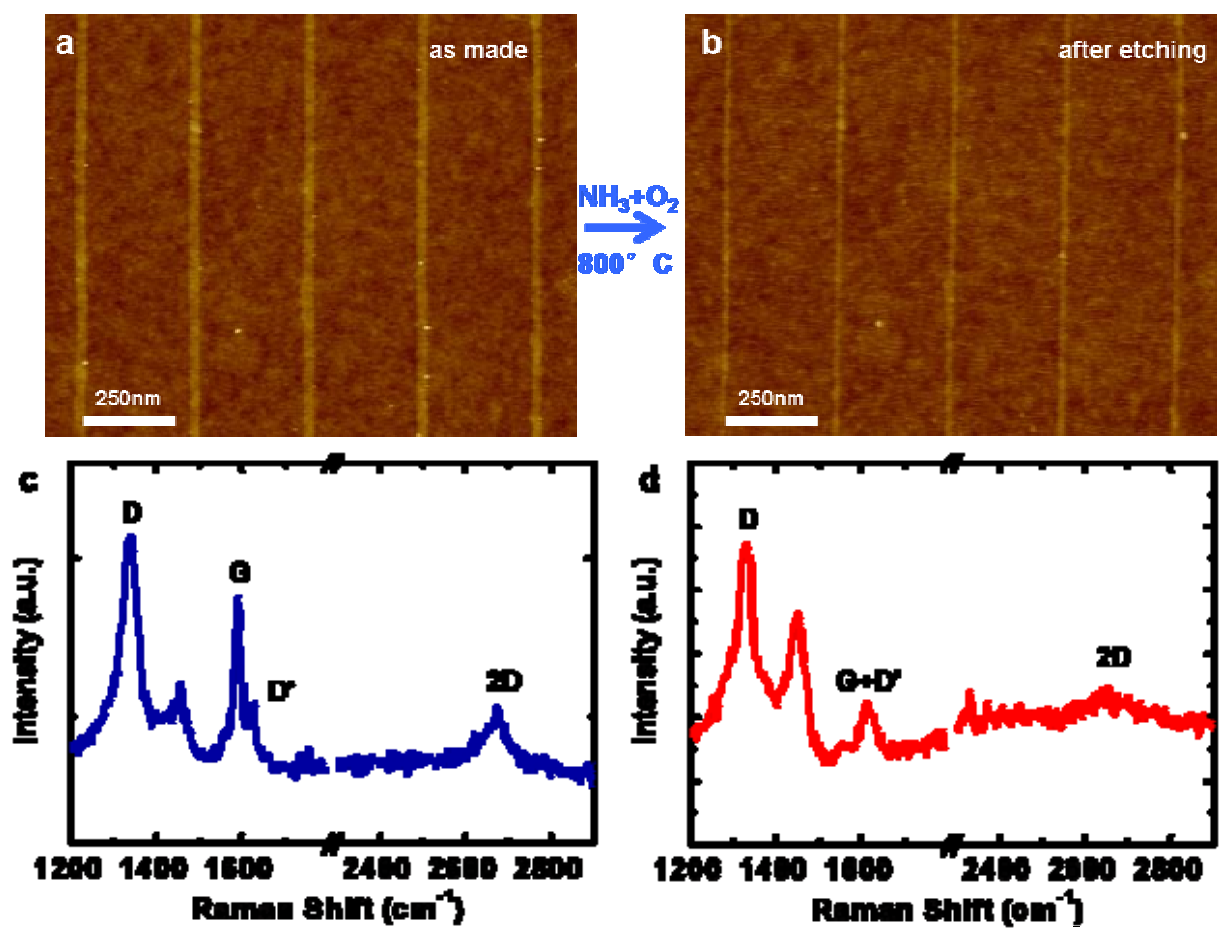

Figure 3

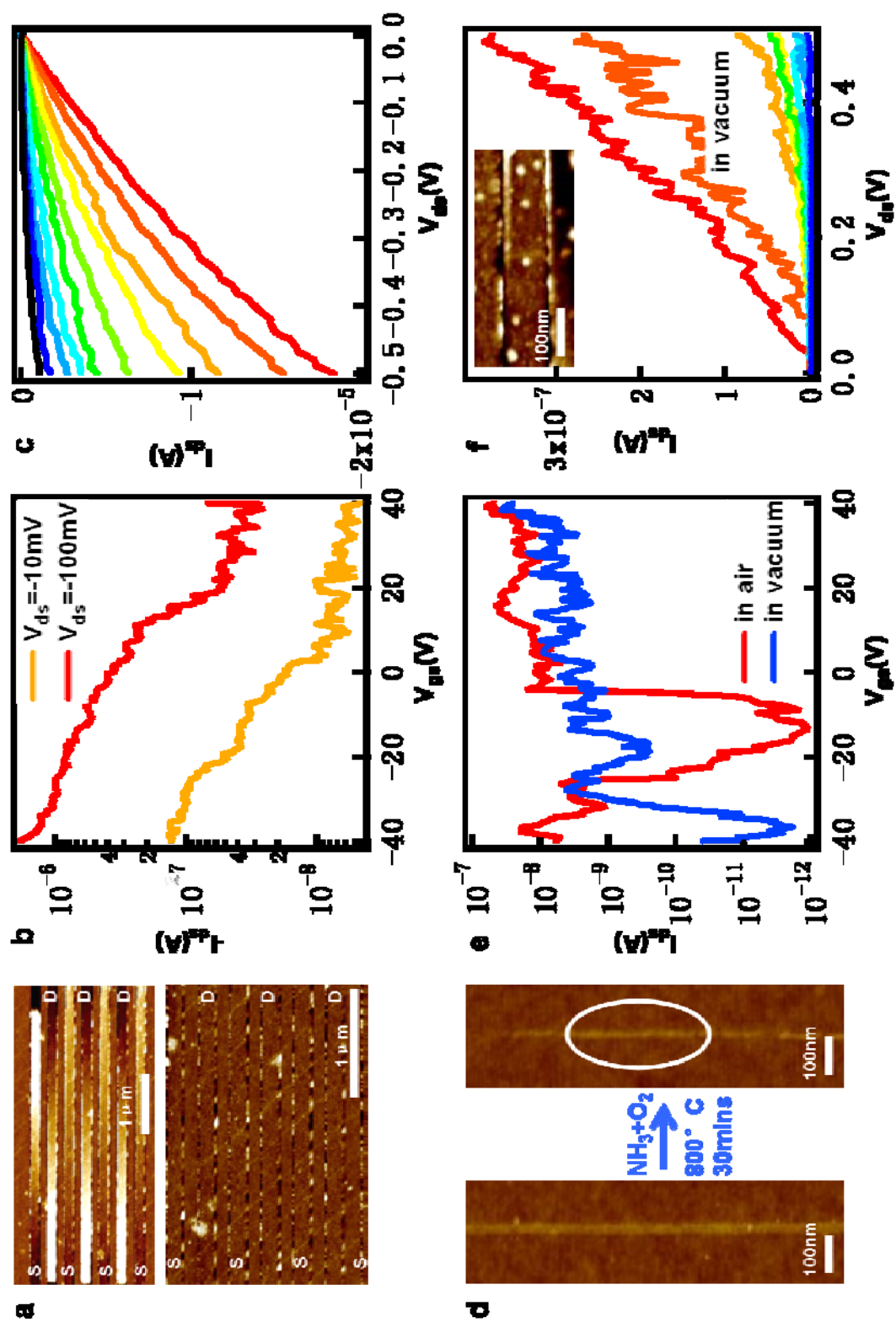

Figure 4